\documentclass[aps,prl,twocolumn,longbibliography,superscriptaddress]{revtex4-2}

\usepackage{amsfonts}
\usepackage{amsmath}
\usepackage{graphicx}
\usepackage{dsfont}
\usepackage{diagbox}
\usepackage{array}
\usepackage{amssymb}
\usepackage{bbm}
\usepackage{float}
\usepackage{subfigure}
\usepackage{MnSymbol}
\usepackage{url}
\usepackage{times}
\usepackage{manfnt}
\usepackage{booktabs}
\usepackage{xcolor}
\definecolor{darkblue}{HTML}{004D6B}
\definecolor{darkred}{HTML}{8c1515}
\definecolor{darkgreen}{HTML}{006400}
\usepackage{hyperref}
\hypersetup{
    pdftitle={MajoranaLiquid},
	colorlinks=true,
	urlcolor=darkred,
	citecolor=darkblue,
	linkcolor=darkred,
	breaklinks
}
\pagecolor{white}
\usepackage{physics}
\usepackage{enumitem}
\usepackage{wasysym}

\begin{document}

\title{Structured volume-law entanglement in an interacting, monitored Majorana spin liquid}

\author{Guo-Yi Zhu}
\email{gzhu@uni-koeln.de}
\affiliation{Institute for Theoretical Physics, University of Cologne, Z\"ulpicher Straße 77, 50937 Cologne, Germany}

\author{Nathanan Tantivasadakarn}
\affiliation{Walter Burke Institute for Theoretical Physics and Department of Physics, California Institute of Technology, Pasadena, CA 91125, USA}

\author{Simon Trebst}
\affiliation{Institute for Theoretical Physics, University of Cologne, Z\"ulpicher Straße 77, 50937 Cologne, Germany}

\date{\today}


\begin{abstract}
Monitored quantum circuits allow for unprecedented dynamical control of many-body entanglement. 
Here we show that random, measurement-only circuits, 
implementing the competition of bond and plaquette couplings of the Kitaev honeycomb model,
give rise to a structured volume-law entangled phase with subleading $L \ln L$ liquid scaling behavior.
This interacting Majorana liquid takes up a highly-symmetric, spherical parameter space within the
entanglement phase diagram obtained when varying the relative coupling probabilities.
The sphere itself is a critical boundary with quantum Lifshitz scaling separating the volume-law phase
from proximate area-law phases, a color code or a toric code.
An exception is a set of tricritical, self-dual points exhibiting effective (1+1)d conformal scaling 
at which the volume-law phase and both area-law phases meet. 
From a quantum information perspective, our results define error thresholds for the color code 
in the presence of projective error and stochastic syndrome measurements.
We show that an alternative realization of our model circuit can be implemented using unitary gates 
plus ancillary single-qubit measurements only.
\end{abstract}
\maketitle


With the advent of digital quantum computing platforms, quantum researchers can now do pioneer work in shaping 
entanglement in quantum many-body systems at will through the implementation of quantum circuits. 
In addition to conventional unitary gates, a decisive element turns out to be the inclusion of non-unitary measurements 
that have been realized to provide an alternative route to the creation of long-range entanglement, 
either in combination with unitaries 
\cite{Raussendorf2001, Raussendorf2005, Raussendorf2006, Stace2016, Piroli21, Verresen2021cat, Tantivasadakarn2021measure, Fisher2022reviewMIPT,Li2018, Skinner2019, Turkeshi22, Hsieh2022, Bravyi2022, NishimoriCat, Lee22decoding, friedman2022locality, Tantivasadakarn2022nonabelian, Tantivasadakarn2022hierarchy, Iqbal2023preparetoriccode}
or even in measurement-only circuits \cite{Barkeshli2021measure, Hsieh2021measure, Khemani2021measure, Buchhold22measurespt,Barkeshli2021measuretoriccode,  Haah21honeycomb, Aasen22, davydova2022floquet, kesselring2022anyon,bombin2023unifying,zhang2022x,vu2023measurement} without any unitary gate evolution.
Instead it is the non-commutativity of the measurement operators that induces entanglement,
which can even exhibit volume-law scaling.

\begin{figure}[b] 
   \centering
   \includegraphics[width=\columnwidth]{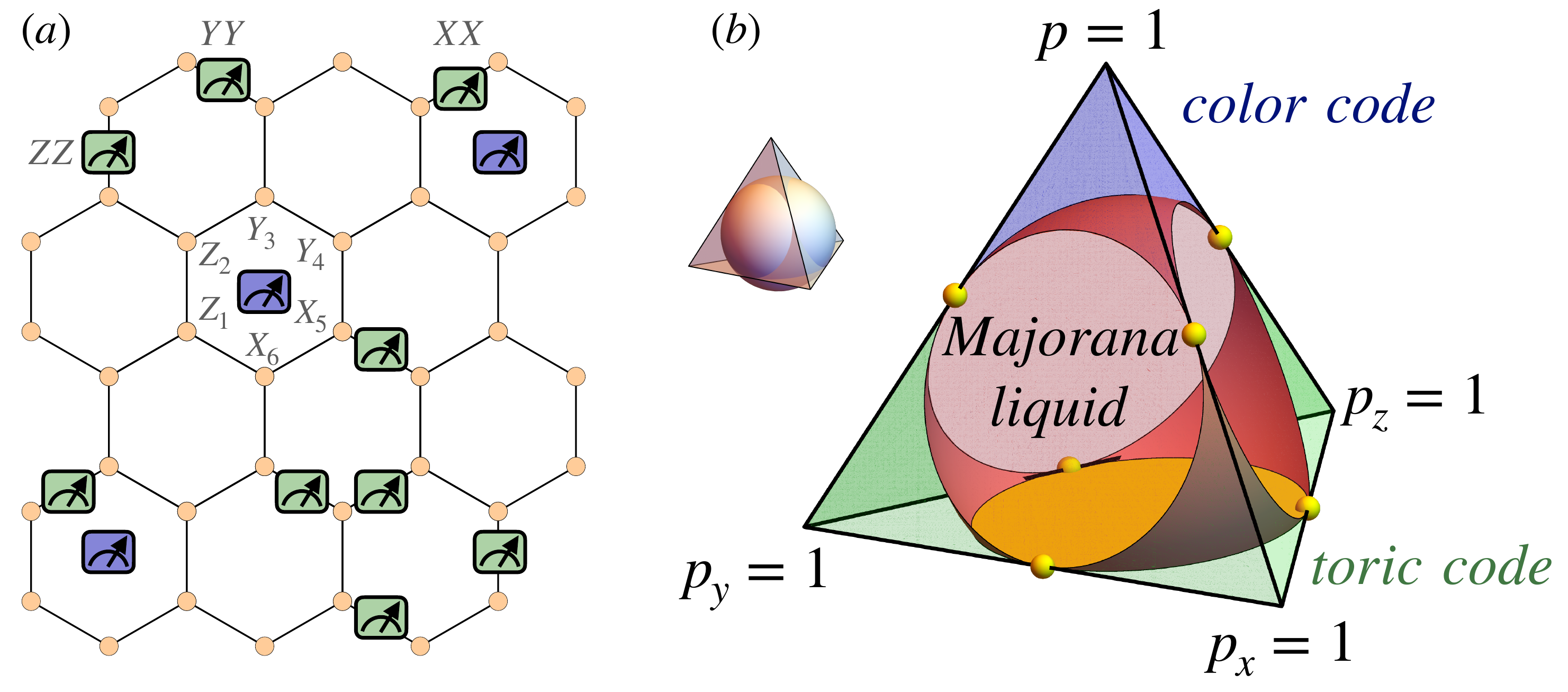} 
   \caption{{\bf Schematics of model and phase diagram}. 
   (a) (2+1)-dimensional random measurement-only circuit on the honeycomb lattice with physical qubits on the sites. Measurements are performed over randomly chosen local bond or plaquette operators, as schematically shown. 
   (b) Schematic quaternary phase diagram drawn as a tetrahedron. 
   A sphere 
   tangent to the edges of the tetrahedron
   cuts the tetrahedron (inset) into four gapped phases separated by a bulk gapless phase.
   The top corner of the tetrahedron stands for the topological color code, while the three bottom corners correspond to the toric code.    The bottom plane of the tetrahedron corresponds to the monitored Kitaev honeycomb model~\cite{Vijay22, Ippoliti22}, i.e.\  a free-fermion limit. 
 {\bf Entanglement structure:}
   The gapless bulk phase enclosed by the sphere is an {\it interacting Majorana liquid} with coexisting volume-law  and $L\ln L$ entanglement scaling. At its boundary (red sphere) it exhibits quantum Lifshitz scaling. .
 The yellow disk at the bottom as well as the yellow (self-dual) dots at the edge centers indicate 
 $L\ln L$ scaling entanglement beyond a pure area-law.
  }
   \label{fig:model}
\end{figure}

In this manuscript, we provide an explicit example of random, measurement-only quantum circuits that induce
{\it structured} volume-law phases in two-dimensional qubit arrays where in addition to an extensive scaling form 
there is an $L \ln L$ scaling, reminiscent of the conformal scaling of quantum liquids with a nodal Fermi surface \cite{Wolf2006,Klich2006}. 
Our model circuit, schematically illustrated in Fig.~\ref{fig:model}(a), 
randomly samples the bond and plaquette couplings of the Kitaev honeycomb model, which can be either represented
as two or six qubit Clifford gates or, alternatively, thought of as Majorana bilinears and a 6-Majorana interaction term.
Crucially, the two types of couplings are not only non-commuting but also stabilize different topological states of matter 
-- a toric code stabilized by the bilinear interactions \cite{Kitaev2006} versus a color code induced by the plaquette interaction \cite{Bombin2006colorcode,Fu2015majoranacode}.
Some of this competition has been previously explored \cite{Vijay22, Ippoliti22} concentrating on the bilinear couplings only,
i.e.\ a monitored circuit analogue of the Kitaev honeycomb model~\cite{Kitaev2006}. There, it was shown that the frustration
of the non-commuting bilinear couplings induce a gapless spin liquid with $L\ln L$ Fermi-surface-like entanglement entropy \cite{Vijay22, Ippoliti22}, contrasting the Majorana Dirac cones of the  Kitaev spin liquid.
Here, we depart the free Majorana fermion scenario by including the additional plaquette coupling, and show that this has a
dramatic effect on the entanglement structure of the many-qubit system. The entanglement phase diagram, illustrated using barycentric coordinates of the probabilities of the four competing terms, is dominated by the emergence of an {\it interacting} Majorana liquid. 
Inside a spherically-bounded phase towards the center of the tetrahedron (Fig.~\ref{fig:model}(b)), we find volume-law scaling of the entanglement entropy 
with an additional $L \ln L$ contribution, inherited from the non-interacting Majorana liquid phase \cite{Vijay22, Ippoliti22} 
inside the circular cut of this sphere with the (non-interacting) base plane of our tetrahedron (marked in yellow in the phase diagram). 
Such a state withstands a structureless thermalized state~\cite{Deutsch91eth, Srednicki94eth} but rather implies the existence of an extensive number of conserved gapless modes like in a Fermi liquid~\cite{Swingle12FLa, Yang12FLent}. We therefore identify this phase with an {\it interacting Majorana liquid}, akin to an interacting Landau-Fermi liquid versus a free-fermion metallic state. 

\begin{figure}[t] 
   \centering
   \includegraphics[width=\columnwidth]{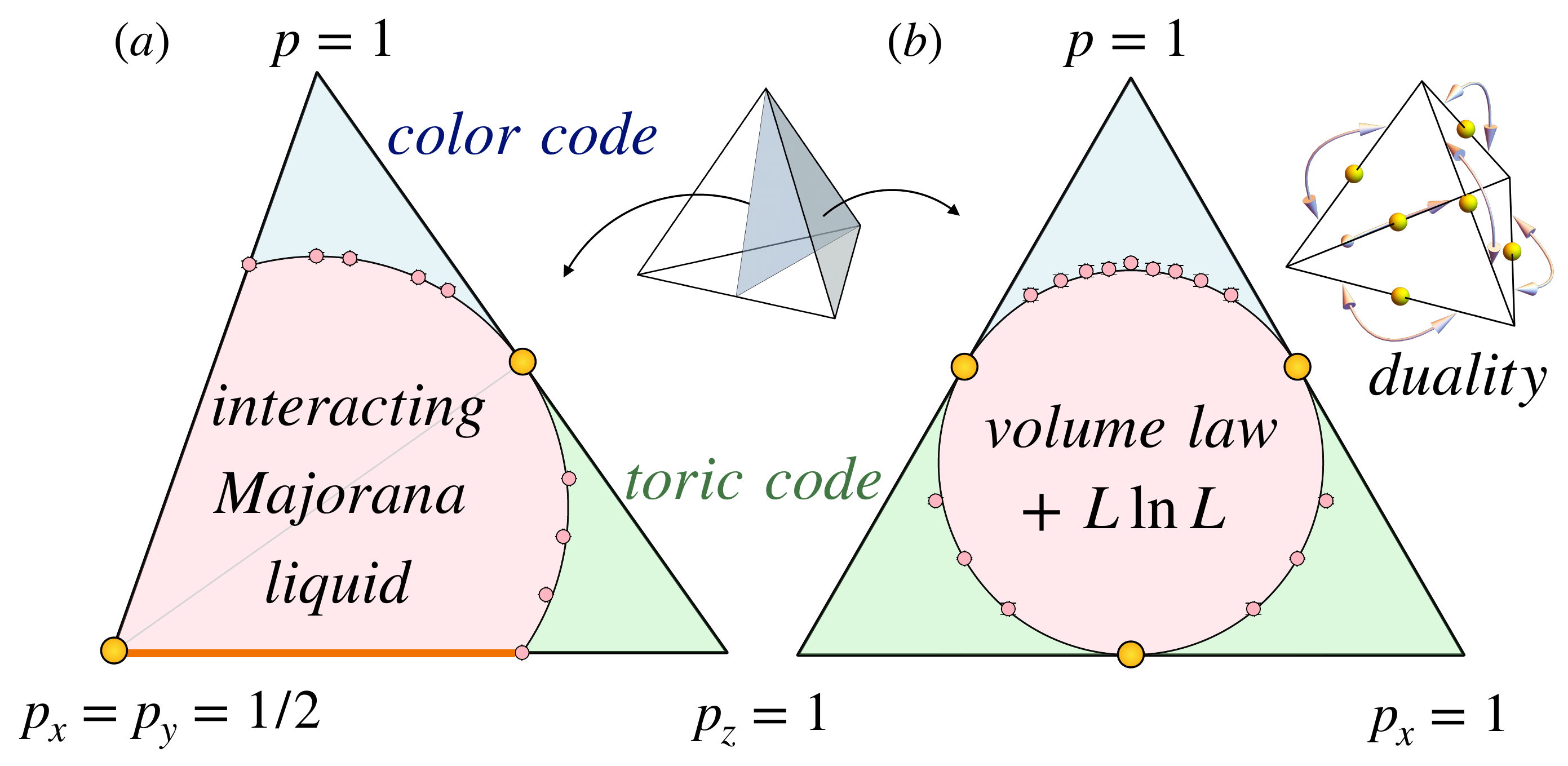} 
   \caption{{\bf Cuts through the tetrahedral phase diagram.}
   Panel (a) shows a middle cut plane, described by $p_x=p_y$,
   (b) is the  side face of the tetrahedron described by $p_y=0$.
   The location of phase transitions (indicated by the pink dots) have been deduced from the finite-size scaling of the tripartite entanglement 
   (see, e.g., Fig.~\ref{fig:I3}(a) below) by sweeping $p$ and $p_z$. 
   The solid line is a sphere tangent to the edge of the tetrahedron.
   The yellow dots indicate self-dual points at the edge centers of the tetrahedron, with the inset on the right illustrating the dualities.
   The bottom orange line indicates the non-interacting Majorana liquid.
   }
   \label{fig:phasediagram}
\end{figure}

The phase boundary of this interacting Majorana liquid, numerically determined in Fig.~\ref{fig:phasediagram}, approximates a perfect sphere tangent to the edges of the tetrahedron. On this spherical boundary we find quantum Lifshitz scaling of 
the entanglement entropy. At the six tangent points, we find a dimensional reduction into stacked (1+1)-dimensional percolation models and a rigorous duality that can flip each edge of the tetrahedron, and thus the six edge centers are self-dual critical points.
Upon perturbation along the edges, they immediately flow to the gapped corner phases of the tetrahedron, while perturbation perpendicular to the edges flow them into the volume law gapless liquid. 
The six solvable edges with their self-dual points pin the global topology of the phase diagram. Nevertheless, the almost perfect spherical geometry of the phase boundary indicates an additional hidden rotation symmetry. 


{\it Model.--}
We consider a random, measurement-only circuit on a honeycomb lattice of size $N=2L^2$, see Fig.~\ref{fig:model}(a). 
In each microstep, we measure a single, randomly chosen Kitaev-type bond-dependent interaction $K=Z_AZ_B,\ (X_AX_B),\ (Y_AY_B)$ with probability $p_{x(y)(z)}$, or alternatively measure the 6-spin interaction $V=Z_1Z_2X_3X_4Y_5Y_6$ with probability $p$. One sweep consists of $L^2$ number of random measurements and will be denoted as one time unit. Note that the operators measured within one sweep do not have to commute with one another, and thus cannot be simultaneously done in one step of the circuit~\cite{NishimoriCat}. 
The non-commuting nature of the measured operators is in fact the crucial ingredient to frustration physics and dynamics~\cite{Khemani2021measure, Vijay22, Ippoliti22}. 
Note also that $V$  is distinct from the conserved Wilson plaquette operator $W=X_1Y_2Z_3X_4Y_5Z_6$ and does {\it not} commute with all the bond checks. 
In a rotated qubit representation, $W$ and $V$ together stabilize a topological color code \cite{Bombin2006colorcode}. 
In the fermion representation~\cite{Kitaev2006}, where each spin is factorized into a Majorana fermion $c_j$ and a gauge field $u_l=\pm1$, $K=iuc_Ac_B$ is the Majorana fermion hopping, and $W=\prod_{l\in \hexagon}u_{l}$ stabilizes the gauge flux, while $V=-iu_{12}u_{34}u_{56}(c_1c_2c_3c_4c_5c_6)$ is the gauged 6-Majorana interaction that stabilizes the Majorana surface code~\cite{Fu2015majoranacode}.

In executing our circuit, we start from an initial flux-free state $\ket{\psi}=\left(\prod_{q}\frac{1+ W_q}{2}\right)\ket{\uparrow}^{\otimes N}$ \footnote{Such states can be prepared by projectively measuring $W_q$ over every plaquette $q$ once upon the spin product state $\ket{\uparrow}^{\otimes N}$ in a finite depth circuit. $\ket{\uparrow}$ is the eigenstate of Pauli $Z$. Note that static flux defects with $W_q=-1$ would result in exactly the same entanglement properties. }.
This initial state we then evolve  until it reaches its steady state, i.e.\ for sufficiently long times of order $O(L)$. 
Since the gauge flux 
is frozen in our circuit model, the ensuing dynamics is solely carried by the Majorana fermions subject to a competition of hopping and plaquette interactions. 
Our model is thus a Clifford stabilizer circuit~\cite{Gottesman04stabilizer} analogue to the ground state of an interacting Majorana Hamiltonian $H\sim (1-p)K + p V$, interpolating between the Kitaev honeycomb model and the Majorana surface code model
\cite{Fu2015majoranacode}. 
While this interacting (2+1)-dimensional lattice Hamiltonian is in general hard to solve, the Clifford stabilizer circuit allows 
for efficient numerical calculation with polynomial scaling, by keeping track of the $N$ generators of the stabilizer group rather than the $2^N$-dimensional quantum many-body wavefunction. 

\begin{figure*}[htbp] 
   \centering
   \includegraphics[width=2\columnwidth]{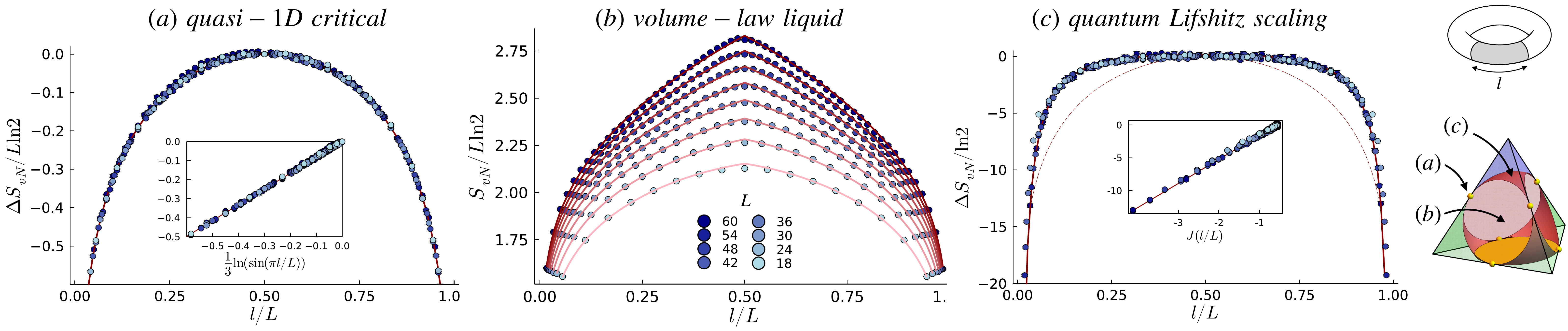} 
   \caption{{\bf Entanglement structure of three forms of gapless matter} in our phase diagram characterized by their entanglement entropy scaling.
   (a) Interacting Majorana gapless state at self-dual point $p=p_z=1/2, p_x=p_y=0$ with dimension reduction. We subtract the area law background to exhibit data collapse for the universal super-area-law correction $\Delta S_{vN} \equiv (S_{vN}(l) - S_{vN}(L/2) )/L= \frac{c}{3}\ln\sin\frac{\pi l}{L}$. We find that $c=0.829(2)\ln2$. 
   (b) Interacting Majorana gapless state exhibits weak volume law scaling with strong $L\ln L$ correction, at the center of the tetrahedron phase diagram $p_x=p_y=p_z=1/4$. 
   The red lines illustrate the fitting scaling function, for $a=1.615(4)\ln2$, $v=0.00951(7)\ln2$, $c=0.642(7)\ln2$, $c'=2.2(2)\ln2$, $\gamma=1.4(1)\ln2$. See Fig.~\ref{fig:SvNdecomp} in SM for a view of each decomposed fractions. 
   (c) Entanglement entropy at the critical point between interacting Majorana liquid and Majorana surface code $p_c=0.683$, $p_x=p_y=p_z$. The solid red line denotes the scaling function with best fit coefficients $\beta=3.67 (3)\ln2$, $\lambda=3.8(2)\ln2$. The dashed line shows the best fit of scaling function $\ln \sin(\pi l/L)$ for comparison, which significantly deviates from the numerical data. The inset shows the data versus rescaled horizontal axis according to the scaling ansatz. 
   }
   \label{fig:SvN}
\end{figure*}


{\it Entanglement phase diagram.--}
The key feature characterizing the dynamically generated, steady-state phases of our monitored quantum circuit is the von Neumann entanglement entropy.
To set the stage, let us first consider the non-interacting set up corresponding to the bottom plane of our tetrahedron.
Here the random bond checks measure the local Majorana fermion parity and effectively teleport single Majorana fermions~\cite{Nayak2008measurement}. 
The final state is a Gaussian fermionic state, a product of long-range Majorana pairs, that exhibits $L\ln L$ Fermi-surface-like entanglement entropy~\cite{Vijay22, Ippoliti22} (see also Fig.~\ref{fig:SvNfreefermioniso} of the supplemental material (SM)). 
By viewing each Majorana pair as a dimer and upon disorder average that, crucially, restores translation symmetry, one can view this non-interacting Majorana liquid as a dynamically generated density matrix analogue of the long-range resonating-valence-bond (RVB) state~\cite{Anderson1973rvb}.
If we now depart the free-fermion setting, an onset of 6-Majorana interaction measurements glues the Majorana pairs beyond the Gaussian fermion state. 
A priori, it is not  clear whether the paired free Majoranas and their consequent $L\ln L$ entanglement can survive this interaction effect. 

To explore this, we analyze the von Neumann entanglement entropy
\footnote{
The entanglement entropy can be efficiently calculated from the number of linearly independent stabilizers supported in a subsystem subtracted by its number of qubits~\cite{Gottesman04stabilizer}, which can be numerically computed by the rank of the stabilizer matrix over the binary field~\cite{Nemo.jl-2017}. 
}
 for a bipartition of the torus (of length $L$-by-$L$) into two cylinders with smooth boundary of fixed length $L$, but varying subsystem bulk length $l$, see the inset of Fig.~\ref{fig:SvN}. 
We consider a most general scaling ansatz of the form
\begin{equation}
S_{vN} (l, L) = 
v \cdot vol(l,L)
+ \frac{cL+c'}{3}\ln\left(\frac{L}{\pi}\sin\frac{\pi l}{L} \right)
+ aL
-\gamma.
\label{eq:SvNscal}
\end{equation}
Here $vol(l, L)=2Ll\ln2 - 2^{4Ll-N-1}$ ($l\leq L/2$) is the volume-law contribution with a leading order Page correction~\cite{Page93},  the second term is a subleading contribution \cite{CalabreseCardy2004, Turkeshi20} that can account for gapless modes akin to a Fermi surface (when viewed as slices of (1+1)-dimensional conformal field theories (CFTs)~\cite{Swingle10fermisurface, Swingle12FLa}). 
We also include an $O(1)$ correction $\gamma$, known as the topological entanglement entropy (TEE)~\cite{KitaevPreskill06tee, LevinWen06tee}. 
 The prefactors $v, c, a$ are non-universal and fitted in our numerics, though we note that $c$ is reminiscent of the central charge in a (1+1)-dimensional CFT.

Let us first consider the case $p=p_z=1/2, p_x=p_y=0$, which is one of the exactly solvable, self-dual points in our phase diagram. 
Coming from the Majorana surface code, the 6-Majorana plaquette interactions stabilize anyon excitations on the plaquettes, while the $ZZ$-bond Majorana bilinear fluctuates these anyons {\it only} along the $z$-direction~\cite{Fu2015majoranacode}. Thus the model is effectively decoupled into stacks of anyon chains and a duality can swap the plaquette interaction and the Majorana bilinear, akin to the Kramers-Wannier duality of the quantum Ising chain~\cite{Kogut79rmp}. 
For further discussion, see the underlying frustration graph given in the SM \footnote{Note that the edge can also be alternatively mapped to a Wen plaquette model~\cite{Wen03plaquette} (i.e.\ a basis-rotated version of the toric code) subjected to uniform $Z$-type measurements.}.  
Each chain can be mapped to a classical 2d bond percolation problem~\cite{Buechler20, Nahum20majorana}, where the prefactor $c$ is exactly calculated  employing CFT 
to be
	$c=3\sqrt{3}\ln 2/(2\pi)$, 
perfectly consistent with our numerical results in Fig.~\ref{fig:SvN}(a). 

Except at these self-dual points, the effect of a non-vanishing Majorana interaction is the immediate formation of a volume-law contribution. As a representative example we show, in Fig.~\ref{fig:SvN}(b), the entanglement entropy for the centroid of the tetrahedral phase diagram, $p_{x(y)(z)}=p=1/4$, the point with qualitatively strongest frustration. The growth of the entanglement entropy with increasing $l$ clearly goes beyond the arc-like $\ln\left(\sin\frac{\pi l}{L}\right)$ scaling of the free fermion limit, but instead an almost linear increase is found for  lengths $l \sim L/2$, resulting in a cusp-like feature known from Page scaling \cite{Page93}. 
Note that even though a volume law is the leading contribution in the $L\to\infty$ (thermodynamic) limit, its prefactor turns out to be {\it two orders of magnitude smaller} than the coefficient of the subleading $L\ln L$ correction, which for small system sizes quantitatively dominates. 
The existence of such an $L\ln L$ correction implies that the volume-law phase is not structureless,
which we further comment on in the discussion section below.
When one moves along the bond-isotropic line $p_{x(y)(z)}=(1-p)/3$ and gradually increases $p$ from 0, the volume-law prefactor rapidly but smoothly grows to a peak value around $p\sim 0.15$ before decreasing again and fading away around $p\sim 0.5$, as shown explicitly in the SM. To diagnose the precise critical point of the transition out of the volume-law phase we resort to the tripartite mutual information ~\cite{Pixley20mutual}. 

At these interacting critical points, the entanglement entropy is found to significantly deviate from the $L\ln L$ correction~\cite{Ju12LlnL} in Eq.~\eqref{eq:SvNscal} and instead exhibits quantum Lifshitz scaling~\cite{Stephan13qlm, Chen15qlm},  originally derived for the gapless dimer RVB state (quantum Lifshitz field theory~\cite{Fradkin04RK})
\begin{equation*}
S_{vN} = aL 
+
\beta J(l/L)+\ldots \,,
\end{equation*}
where $J(x)=-\ln\frac{\theta_3(i\lambda x)\theta_3(i\lambda(1-x))}{\eta(2ix)\eta(2i(1-x))} $, with $\theta_3$ the Jacobi-theta function and $\eta$  the Dedekind-eta function~\cite{Stephan13qlm, Chen15qlm}, while $\lambda$ is a parameter related to the inverse stiffness or correlation exponent  \cite{Fradkin04RK} in the original derivation \footnote{ 
$J(x)$ is derived by mapping the exact dimer wavefunction into the 2-dimensional free boson CFT and tracing out the boson for its partition function~\cite{Stephan13qlm}. 
Note that at the limit $x\ll 1$, the area law correction $J(x)\sim -\frac{\pi}{24x}$, in contrast to $\ln\sin(x\pi)\sim \ln x$ for the (1+1)d CFT, and their respective prefactors $\beta$ and $c$ take qualitatively similar meaning as universal characterization of entangled degrees of freedom. }.
An example of such quantum Lifshitz scaling is shown in  Fig.~\ref{fig:SvN}(c).
On a speculative note, this Lifshitz scaling might be a harbinger of space-time anisotropy with a dynamical critical exponent $z=2$ (though counter-examples \cite{Chen15qlm} indicate that no such stringent connection can be made), 
which would possibly allow us to connect this scaling form to the Lifshitz transition of Fermi surface topologies \cite{Volovik2017} -- an appealing completion to our
scenario of a sequence of transitions from non-interacting to interacting to vanishing Fermi liquid as one ascends the vertical direction in our tetrahedral phase diagram.

\begin{figure}[t] 
   \centering
   \includegraphics[width=\columnwidth]{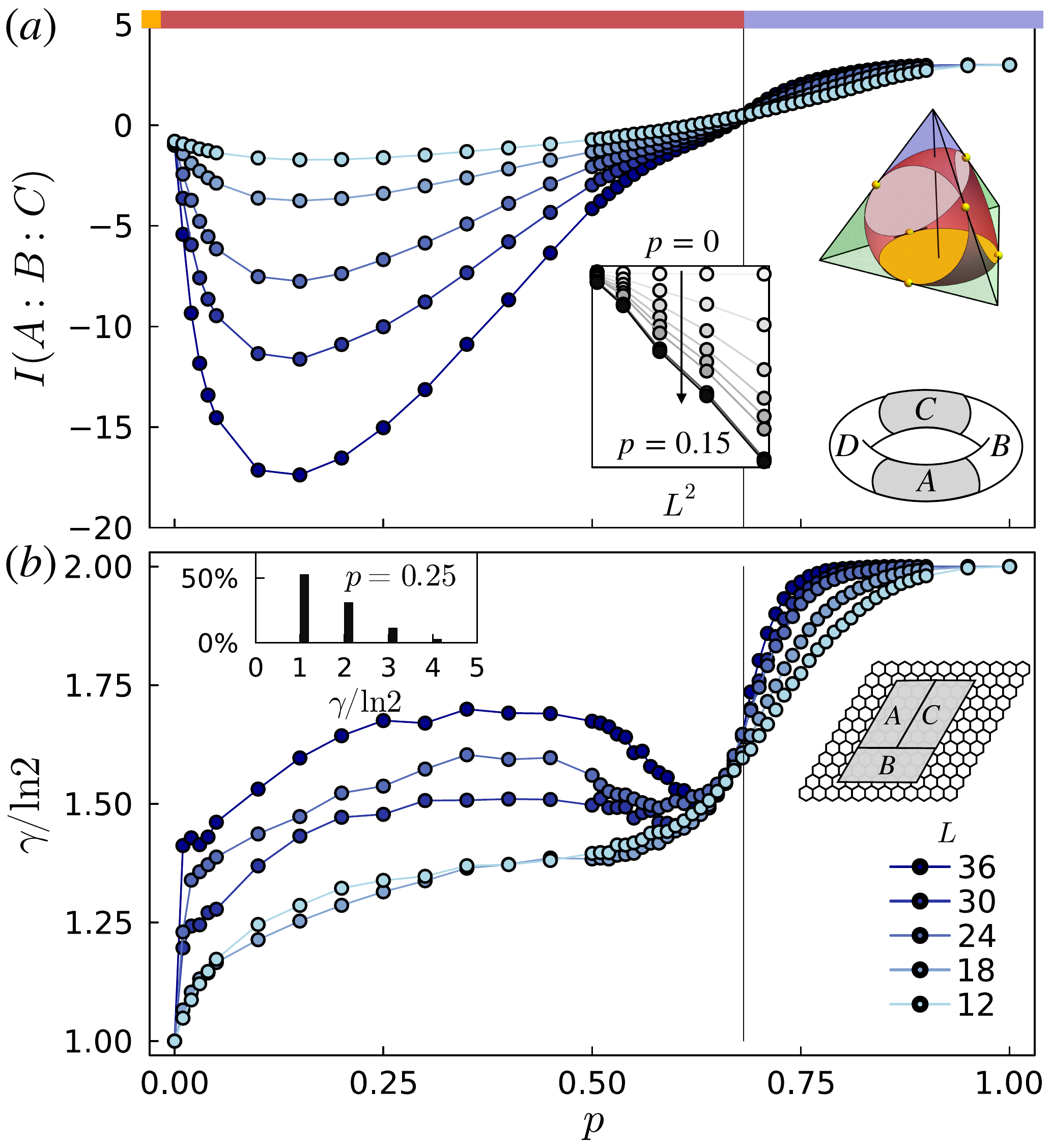} 
   \caption{{\bf Topological and entanglement phase transition} along the bond isotropic line: $p_{x(y)(z)}=(1-p)/3$ (central pillar in the tetrahedron is shown in the inset). The top red bar illustrates the liquid phase while the blue bar illustrates the gapped color code phase, with the yellow dot being the free fermion liquid state. 
   (a) {\bf Tripartite mutual information} between three cylinders (schematically shown in the inset). 
   From finite-size scaling we find $p_c=0.682(4)$, very close to the exact boundary of the sphere $p_c = (1+\sqrt{3})/4 = 0.68301\ldots$, and $1/\nu=1.01(6)$. 
   For the phase region between $p\in(0,p_c)$, the mutual information diverges $I \propto -L^2$ with system size, as shown in the inset for the window $p\leq 0.15$. 
   (b)
   {\bf Topological entanglement entropy}. 
   Inset shows  the distribution of $\gamma$ among the disorder ensemble for $p=0.25$. 
   Data is averaged over 10,000 disorder realizations for $L<30$ and 5,000 samples for $L\geq 30$. 
   }
   \label{fig:I3}
\end{figure}

 {\it Topological codes and phase transitions.-- }
Let us round off our discussion of the entanglement phase diagram by looking at the four corner phases, which are gapped area-law phases realizing either a toric code (for the three bottom corners) or a color code (near the top of our tetrahedron). Starting from one of these gapped phases, we can discuss the entanglement transition into the interacting Majorana liquid. Mapping out the phase boundary can be done, as before, by computing the tripartite mutual information (TMI) 
\begin{equation*}
I(A:B:C)= S_A+S_B+S_C-S_{AB} - S_{BC} - S_{AC} + S_{ABC}
\end{equation*} 
for a partition of the torus into four cylinders (inset of Fig.~\ref{fig:I3}a).
As shown in Fig.~\ref{fig:I3}(a), away from the free fermion limit $p=0$ where $I=-1$~\cite{Vijay22, Ippoliti22}, the TMI is {\it extensive} for the interacting liquid phase, i.e. $I(A:B:C)\propto -L^2$ as shown in the inset. Such an indicator of information scrambling~\cite{Yoshida16I3} is consistent with the volume-law entanglement entropy we found earlier. In the color code limit $p\to 1$, $I(A:B:C)=+3$ due to three independent effective Bell pairs between $A$ and $C$, formed by the product of plaquettes of the color code (its plaquettes being 3-colorable when $L\mod 3=0$). In between, the crossing point indicates an entanglement phase transition from the interacting Majorana liquid to the Majorana surface code, which we used to quantitatively map out the phase diagram of Fig.~\ref{fig:phasediagram}. 

From a quantum information perspective, we can interpret the area- to volume-law transition out of the color code as an {\it error threshold}
 for the color code subject to projective bond errors and stochastic syndrome measurements. 
This is best revealed in the topological entanglement entropy (TEE)~\cite{KitaevPreskill06tee, LevinWen06tee}, calculated for the 
tripartite geometry in the inset of Fig.~\ref{fig:I3}(b). In the color code phase, it shows a plateau at $2\ln 2$, reflecting the two bits of information contributed from the gauge {\it and} Majorana sector (versus one bit in the toric code where only the gauge sector contributes). At the threshold $p_c$ of the color code, the TEE drops from its plateau value signaling the breakdown of topological order.
This transition gives a fundamental upper-bound of the decoding threshold for the color code under such noise. 
The TEE is non-quantized in the interacting liquid regime (while still showing a system size dependence, growing with increasing $L$).
We note that the volume-law phase can still be used as a code space with quantum error correction \cite{Altman2020qec, Huse2020qec}, 
but (in light of the small volume-law prefactor) it might be much less effective in storing logical quantum information, 
see SM for purification dynamics indicating a corruption of the code space. 

{\it Outlook.-- }
On a technical level, one might wonder whether our highly symmetric phase diagram allows for an analytical understanding. One step in this direction is to pursue a coupled-wire approach: Start from a bottom edge of the tetrahedral phase diagram, which corresponds to stacked monitored Majorana chains, and then turn on either the Majorana hopping or Majorana interactions. The former coupling leads to the free fermion liquid within the bottom plane, while the latter sets off a flow to the volume-law liquid in the side plane of the tetrahedron. Despite this distinction, both directions show surprisingly similar geometrical phase boundaries: a circle, see Fig.~\ref{fig:phasediagram}. This might be related to the similarity of their frustration graph structure, which -- albeit not exactly identical in their microscopic details --  can both be viewed as a stack of bipartite horizontal chains, with interchain degree-4 nodes relating the two sublattices (see SM). 

Concerning the practical implementation of our circuit model, which we have here formulated as a measurement-only, multi-qubit circuit, we note that it can alternatively be implemented by a unitary circuit with two-qubit gates and single-qubit measurements only. To do so, one needs to introduce a set of ancillae qubits~\cite{NishimoriCat}, one for each bond coupling expanding the lattice geometry to heavy-hexagon and one ancilla for each hexagon (see SM for details). Such an implementation is relatively close to current quantum processor designs (such as IBM's transmon platform) and akin to syndrome measurements of the surface code~\cite{Google2023surfacecode}.


{\it Discussion.-- }
A hallmark of equilibrium quantum states of matter is their boundary-law entanglement scaling \cite{Eisert2010},
which for Fermi liquids experiences a mild violation in terms of an $L \ln L$ ``super-area-law" contribution \cite{Wolf2006,Klich2006, Swingle12FLa, Yang12FLent}.
In contrast, the non-equilibrium Fermi liquid discussed in our work exhibits an extensive (volume-law scaling) entanglement entropy,
with a subleading $L \ln L$ contribution in (2+1) dimensions. 
The existence of this subleading term not only distinguishes our state from a conventional {\it thermal} steady state, as postulated by
the eigenstate thermalization hypothesis, but it might prove to be essential:
In its (1+1)-dimensional analogues, it has been argued that the subleading $\ln L$ correction indicates a {\it protection mechanism} of
the volume-law entanglement structure as it originates from a power-law distribution of stabilizers \cite{Li2019} that counteract the detrimental effects of local projective measurements on long-range stabilizers.
One might argue that a similar mechanism plays out in (2+1)-dimensional quantum liquids indicating an essential role for 
the $L \ln L$ term to allow for a {\it stable} volume-law phase as we have observed it in the monitored quantum circuit model at hand
\footnote{It remains an interesting question to explore whether this entanglement structure remains stable when moving away from Clifford stabilizer circuits.}.

The coexistence of volume-law and $L\ln L$ scaling we report here might bear some resemblance with the observation of {\it quantum many-body scars} \cite{serbyn2021scarreview, chandran2022scarreview} in (1+1)-dimensional 
models. 
There one observes a {\it weak ergodicity breaking} that manifests itself in a tower of $\ln L$ entangled non-thermal eigenstates~\cite{Bernevig18akltscar} 
coexisting with the otherwise volume-law entangled thermal states~\cite{Deutsch91eth, Srednicki94eth}.
Instead of starting from an ergodic phase our model arrives at a similar entanglement structure, in a (2+1)d generalization,  from a proximate (super) area-law phase, i.e.\ it exhibits {\it weak information scrambling}. On a  speculative note, this scrambling transition from the non-interacting to interacting Majorana liquid has a renormalization group flavor to it, which manifests itself, e.g., in the sudden change of the $L \ln L$ prefactor, reminiscent of the flow of central charges dictated by the $c$-theorem in (1+1)d CFTs.

A characteristic of our model is its  {\it randomness}, manifest in the space-time disorder of the circuit, in addition to 
measurement outcomes, which results in an {\it ensemble of disordered pure states}. 
This randomness spoils translation symmetries for each individual disorder realization (of the circuit), 
which makes it possible to have a stable Majorana Fermi-surface even in the presence of time-reversal symmetry~\cite{Chua2011}. 
The disorder average then restores the symmetries on a statistical level. 
The disorder averaged entanglement represents {\it typical pure wavefunctions} in the ensemble, but {\it not} the average density matrix, which may be also interpreted as a translationally invariant state in the double Hilbert space~\cite{bao2023mixedstate}.
An interesting future direction is to further explore the essential role of randomness here, e.g.\ by imposing space or time translation symmetry into the protocol \cite{Gullans22crystalline}, such as a spatially random Floquet circuit \cite{Chalker18randomfloquet},  a quasi-periodic protocol~\cite{Potter18quasiperiodic}, or a translationally invariant Floquet protocol with weak measurements ~\cite{ZhuTrebst23honeyweakmeas}. 

Let us close our discussion with a comment on computational complexity.
Highly non-trivial entanglement structures can arise from the competition of local interactions 
-- either in the steady-state of the long-time evolution of a random measurement circuit, as discussed in this manuscript,
or in the quantum ground state of a quantum many-body system cooled down under Hamiltonian dynamics.
Despite their similar ingredients the two approaches come with very different simulation costs on a classical computer
-- an interacting ground state with a Fermi surface is known to create a sign problem~\cite{grossman2023unavoidable}
in quantum Monte Carlo simulations \cite{TroyerWiese2005}, while we have shown here that a similarly entangled state can be simulated with Clifford stabilizer circuits in polynomial time
~\cite{GottesmanKnill98}. 
This leaves us in the fascinating situation that going to the Clifford circuit analogue state 
has {\it reduced} the computational 
complexity of simulating an interacting Fermi liquid -- a route that should be further explored, for other quantum states of interest, in the future.

\begin{acknowledgments}
\textit{Acknowledgments.--} 
We thank Michael Buchhold and Xhek Turkeshi for insightful discussions. 
The Cologne group was partially funded by the Deutsche Forschungsgemeinschaft under Germany's Excellence Strategy -- Cluster of Excellence Matter and Light for Quantum Computing (ML4Q) EXC 2004/1 -- 390534769 and within the CRC network TR 183 (Project Grant No. 277101999) as part of projects A04 and B01.
NT is supported by the Walter Burke Institute for Theoretical Physics at Caltech.
The numerical simulations were performed on the JUWELS cluster at the Forschungszentrum Juelich.
\end{acknowledgments}

\bibliography{measurements}

\clearpage
\appendix

\section{Supplemental Material}

To complement our discussion in the main text, this appendix provides a short discussion of the frustration graph underlying our 
circuit model, further details on an implementation using unitary gates and single-qubit measurements only, as well as supplementary
numerical data of the entanglement structure, particularly of the structured volume-law phase, as well as a short discussion of the purification dynamics in the various phases of our model.

\subsection{Frustration graph}

\begin{figure}[H] 
   \centering
   \includegraphics[width=\columnwidth]{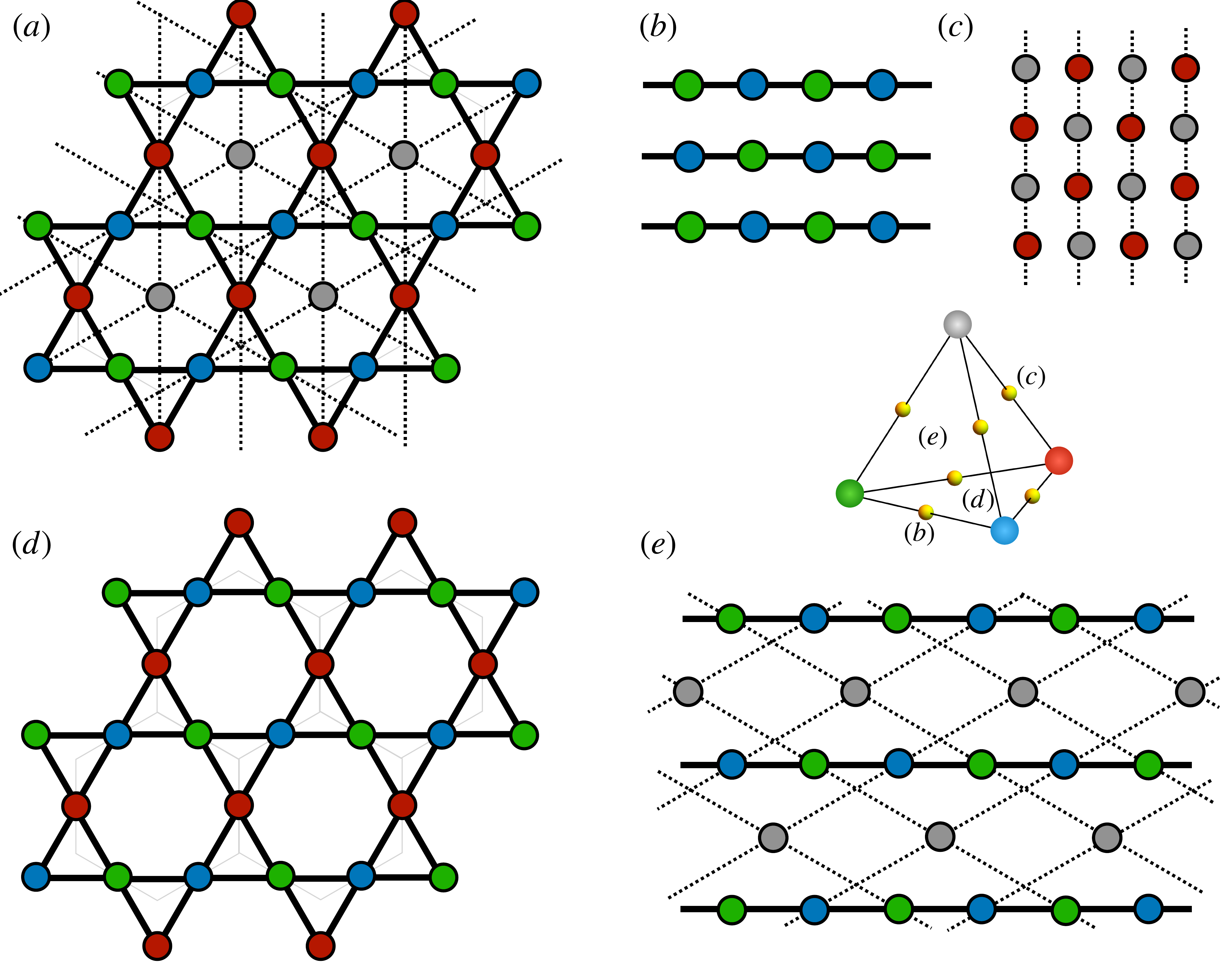} 
   \caption{{\bf Frustration graph, duality and dimension reduction}. 
   Each node stands for an operator being measured. Two operators anticommute (commute) with each other if there is (no) link connecting them. 
   (a) Complete graph for all four types of measurement operators, with blue, green, red nodes representing the bond checks, and the gray node for the plaquette interaction. 
   (b)(c) Reduced graph along the edges of the tetrahedron phase diagram. 
   (d) Reduced graph at the bottom plane of the tetrahedron, forming a Kagome lattice. 
   (e) Reduced graph at the side face of the tetrahedron. }
   \label{fig:graph}
\end{figure}

To understand the dynamics and frustration of the measurement-only protocol, we draw the frustration graph~\cite{Khemani2021measure} as in Fig.~\ref{fig:graph}. The nodes of the graph do not correspond to a basis of the wavefunction, but rather a Pauli operator, which are the ones being measured in our protocol. They live on the dual lattice of the original honeycomb lattice, where the blue, green, red nodes correspond to the bonds of the honeycomb lattice, and the gray nodes correspond to the hexagon plaquette. The probability vector $(p, p_x, p_y, p_z)$ actually determines the ``fugacity" of these nodes i.e. probabilities that they are measured at a given spacetime point. One may view a measurement event as ``occupying" the node, which is nearest-neighbour exclusive in space. The center of the phase diagram $p=p_x=p_y=p_z$ features equal fugacity of all the nodes. 
Along the edges of the tetrahedron, the graph reduces to stacks of 1d chains that has a bipartite structure, connecting only two colors (Fig.~\ref{fig:graph}bc). There is a duality for the dynamics, described in the graph, under swapping the two colors or translation along the chains. Since the real-space support of nodes of different colors are different, the entanglement entropy does not have to be invariant under the color swapping duality. 
%
The bottom plane and the side face of the tetrahedron share a similar graph (comparing Fig.~\ref{fig:graph}de): rows of bipartite translational invariant chains, being coupled by degree-4 interchain nodes. The interchain nodes couple only nearest neighbour for the former case while 3rd nearest neighbour for the latter scenario.

\subsection{Alternative circuit implementation}

\begin{figure}[H] 
   \centering
   \includegraphics[width=\columnwidth]{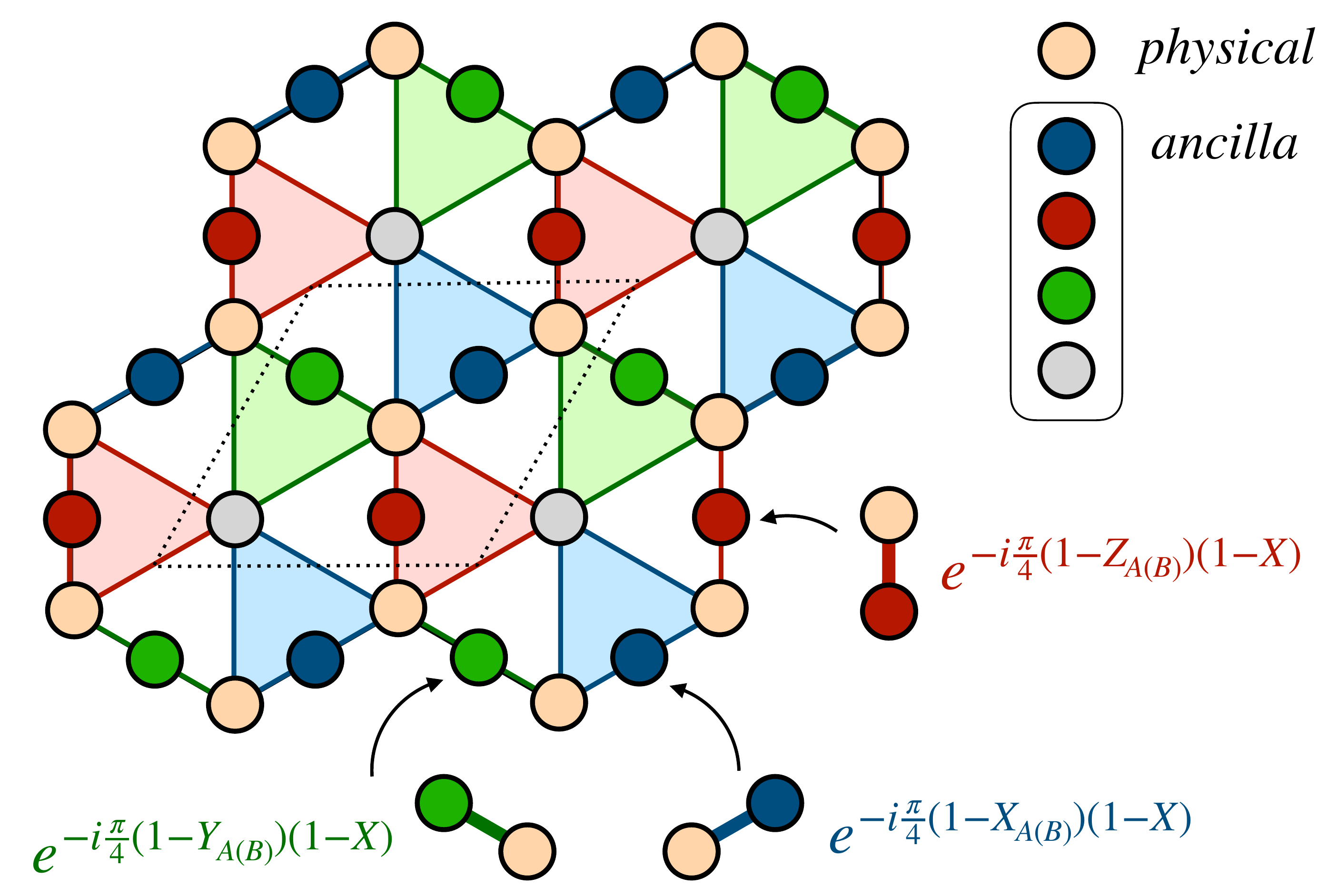} 
   \caption{{\bf Monitored unitary circuit with 1-qubit measurements.}
   	Alternative implementation of our circuit model where the bilinear and 6-spin measurements are implemented via projective measurements of ancilla qubits. 
	The original lattice geometry of qubits (spins) on the hexgonal lattice (orange circles) is thereby expanded to a heavy-hexagon 
	geometry (as implemented, e.g., in the current transmon quantum processors of IBM) with an additional ancilla qubit in the hexagons.
	The dotted line encloses one unit-cell.}
   \label{fig:unitary}
\end{figure}

Let us devise an alternative circuit implementation that removes the need for multiqubit measurement operations. 
To do so, we need to introduce an extensive number of ancilla qubits -- one for each physical operator that we wish to measure.
Namely, we place ancilla qubits on all the bond centers and the plaquette centers of the original honeycomb lattice, see Fig.~\ref{fig:unitary}. In order to couple a given ancilla qubit to our targeted measurement operator, we just need to apply a basis rotated variant of the CNOT gate: $\exp(-i \frac{\pi}{4}(1-Z)(1-X))$, or $\exp(-i \frac{\pi}{4}(1-X)(1-X))$, or $\exp(-i \frac{\pi}{4}(1-Y)(1-X))$, depending on the type the bond and where we always use the physical qubit as the control qubit. All the bonds connecting an ancilla qubit have to be turned on, and since they commute with each other the gate sequence is irrelevant. After that, the ancilla qubit is maximally entangled to the targeted measurement operator and can be measured. 
%
To summarize, such an implementation is in similar spirit as the syndrome measurements of the surface code.

\subsection{Supplemental data}

\subsubsection*{Free Majoranas}

\begin{figure}[H] 
   \centering
   \includegraphics[width=.9\columnwidth]{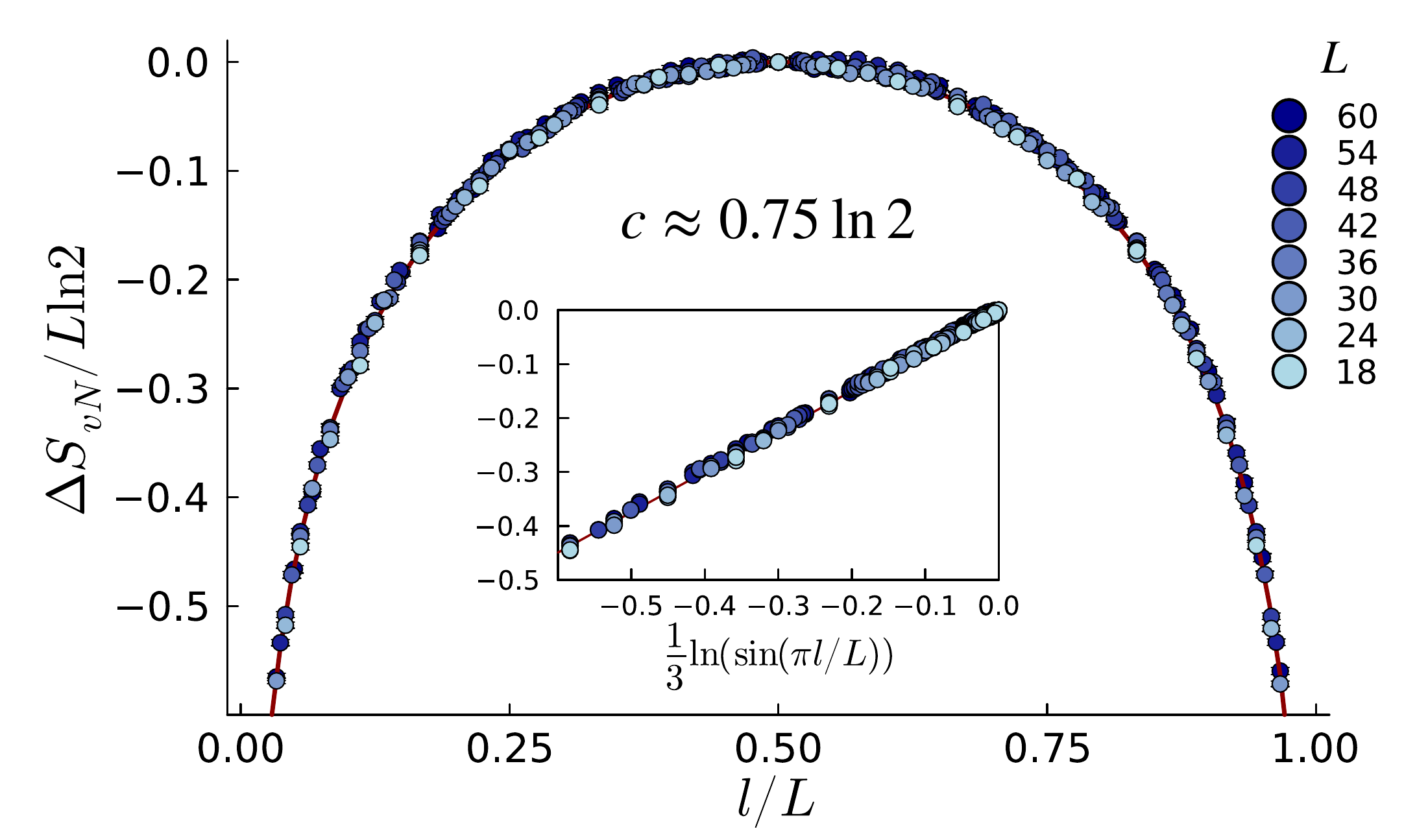} 
   \caption{
   {\bf Entanglement entropy scaling for isotropic free Majorana point} for $p_x=p_y=p_z=1/3$, $p=0$.
	}
   \label{fig:SvNfreefermioniso}
\end{figure}

We complement the entanglement characterization of Fig.~\ref{fig:SvN} in the main text with an additional plot for the free Majorana case in Fig.~\ref{fig:SvNfreefermioniso} showing the entanglement arc for the isotropic point in the base plane of our tetrahedron ($p_x=p_y=p_z=1/3$, $p=0$). The calculation is performed over evolution times $T=50$ for $L<40$ while $T=80$ for $L>40$, averaged over 1000 disorder samples for $L<60$ and 360 samples for $L=60$. 

\subsubsection*{Structured volume-law phase}

\begin{figure}[t!] 
   \centering
   \includegraphics[width=\columnwidth]{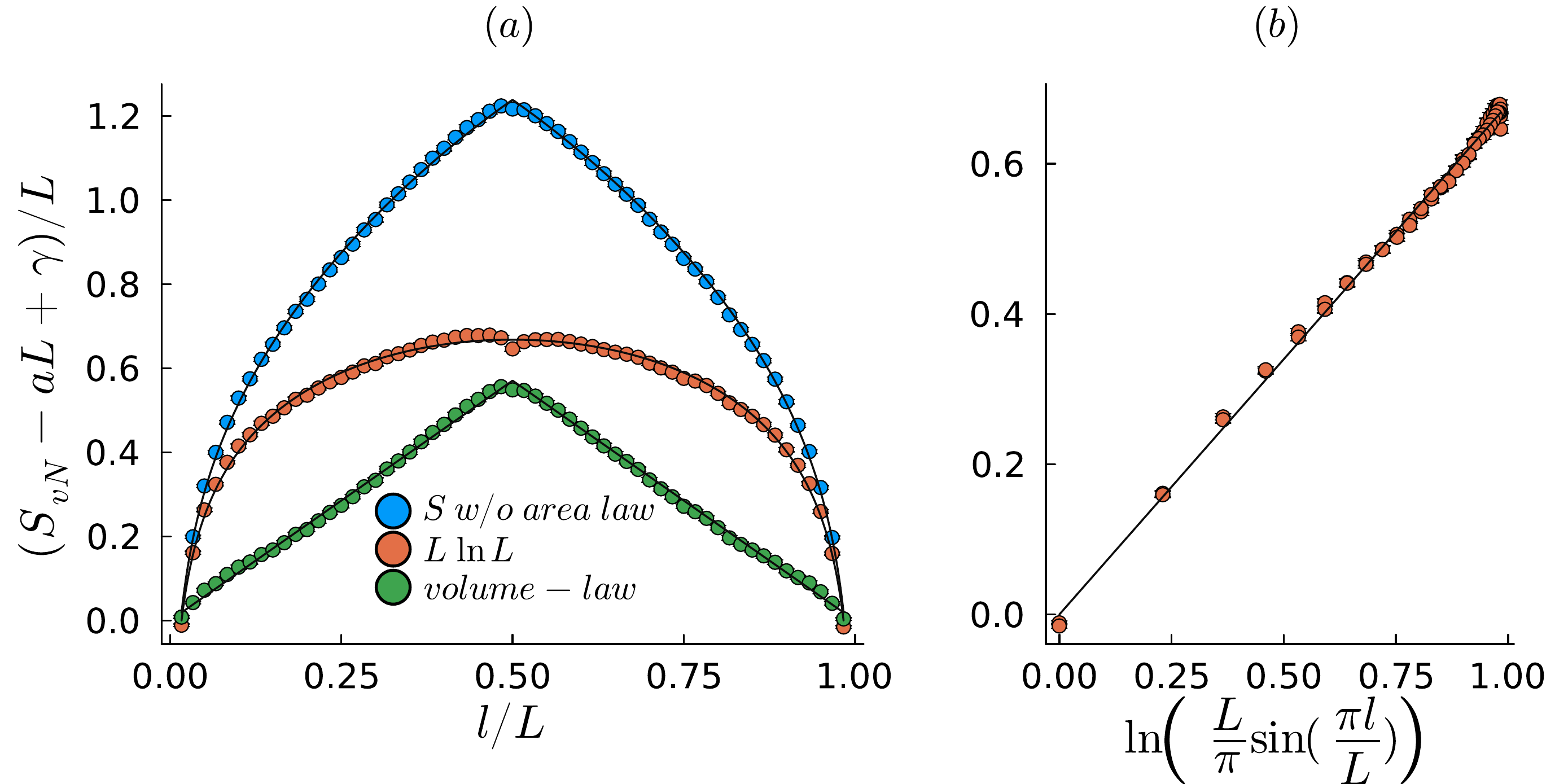} 
   \caption{
   {\bf Decomposed fractions of the entanglement entropy in the interacting Majorana liquid phase}. All solid lines are the best fit scaling function. Data is obtained from $p=p_{x(y)(z)}=1/4$, $L=60$. }
   \label{fig:SvNdecomp}
\end{figure}

For a better view of the three contributing fractions in the volume-law phase according to the scaling ansatz Eq.~\eqref{eq:SvNscal}, we subtract the numerical data of entanglement entropy by the area-law scaling function $aL-\gamma$, and further subtract it by either the fit volume law scaling function $vol(l,L)$, or the gapless scaling function $(cL+c')\ln\left(\frac{L}{\pi}\sin( \frac{l\pi}{L})\right)$, see Fig.~\ref{fig:SvNdecomp} left panel. To see the gapless fraction more clearly, we show it under the rescaled horizontal axis as shown in the right panel, which is approximately linear with slope $(cL+c')/3$. 

\begin{figure}[H] 
   \centering
   \includegraphics[width=.99\columnwidth]{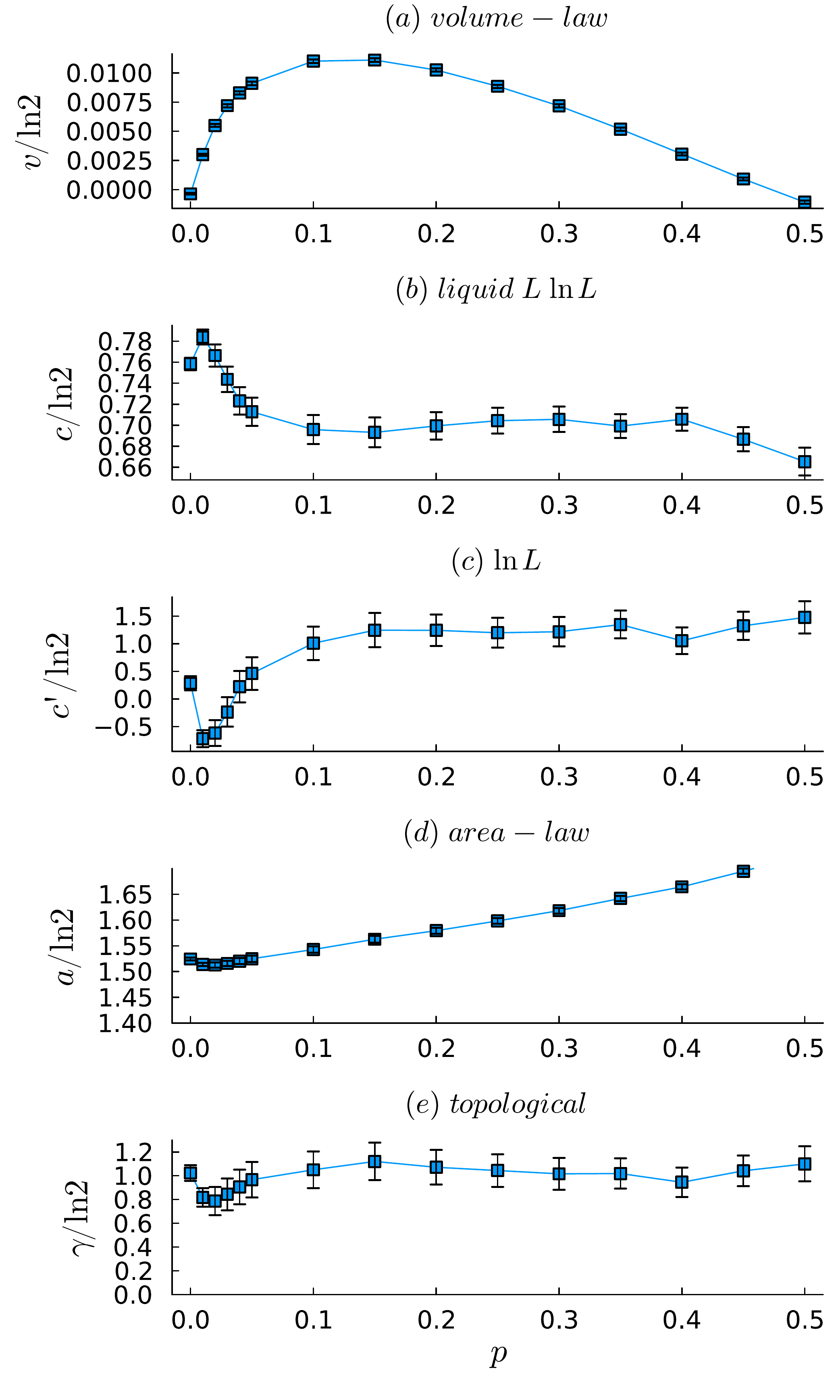} 
   \caption{
   {\bf Evolution of structured volume-law entanglement.}
    Shown are the fitting parameters for the volume-law, liquid, and area law contributions, for a vertical cut through the tetrahedral phase diagram, i.e. $p_x = p_y = p_z$ and $p \in [0,0.5]$. Data is obtained by system size $L=36$ with averaging over 1,000 disorder samples. 
   }
   \label{fig:abc}
\end{figure}

In the volume-law phase, we show the fit coefficients evolving along the isotropic line of the phase diagram by varying $p$ in Fig.~\ref{fig:abc}. The fit is done with system sizes $L=18,24,30,36$. 
Note that the fit coefficient at $p=0.25$ found by $L\leq 36$ is slightly different from the coefficients fit with $L\leq 60$ as shown in the main text,
which we attribute to a finite size drift and the fitting error (note the error bar in fitting $c$ compared with fitting $a$, under the background of volume law and area law contribution). 
When $p>0.5$, the volume-law coefficient is approximately zero, beyond such scaling function. 
At the free fermion limit $p=0.0$, without volume-law entanglement, its entanglement entropy is dominated by the gapless law, which we fit in Fig.~\ref{fig:SvNfreefermioniso} with system sizes up to $L=60$, where the data collapse works perfectly and the prefactor is consistent with Ref.~\cite{Vijay22, Ippoliti22}. 

\subsubsection*{Topological entanglement entropy}

\begin{figure}[H] 
   \centering
   \includegraphics[width=\columnwidth]{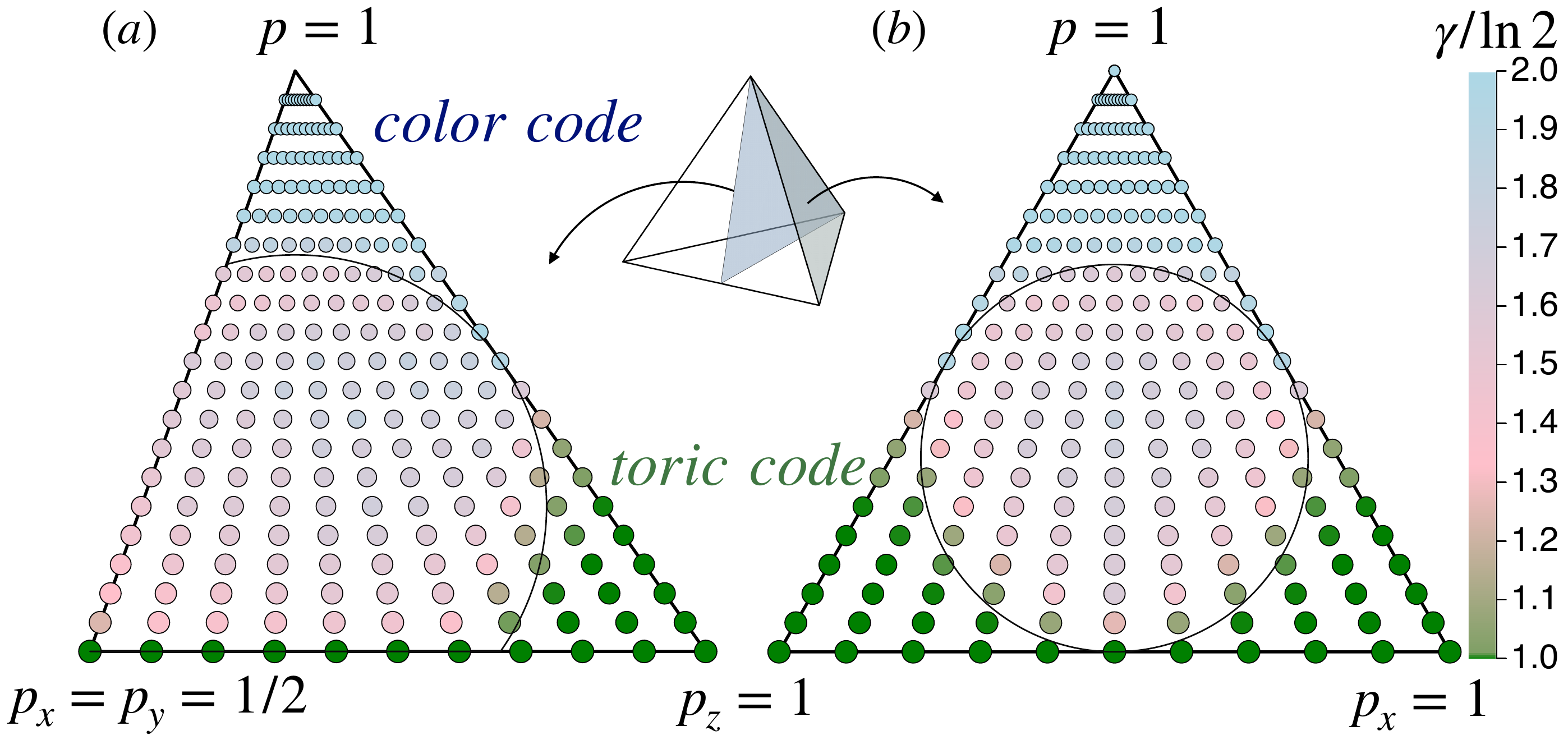} 
   \caption{
  {\bf Topological entanglement entropy} mapped out in the phase diagram, for $L=36$ and averaged over $1,000$ samples. }
   \label{fig:TEEglobal}
\end{figure}

Complementing Fig.~\ref{fig:I3} in the main text, we provide a global view of the topological entanglement entropy in the phase dia\-gram (Fig.~\ref{fig:TEEglobal}). The color coding qualitatively reflects the three topologically distinct phases. 

\subsubsection*{Purification dynamics}

\begin{figure}[H] 
   \centering
   \includegraphics[width=\columnwidth]{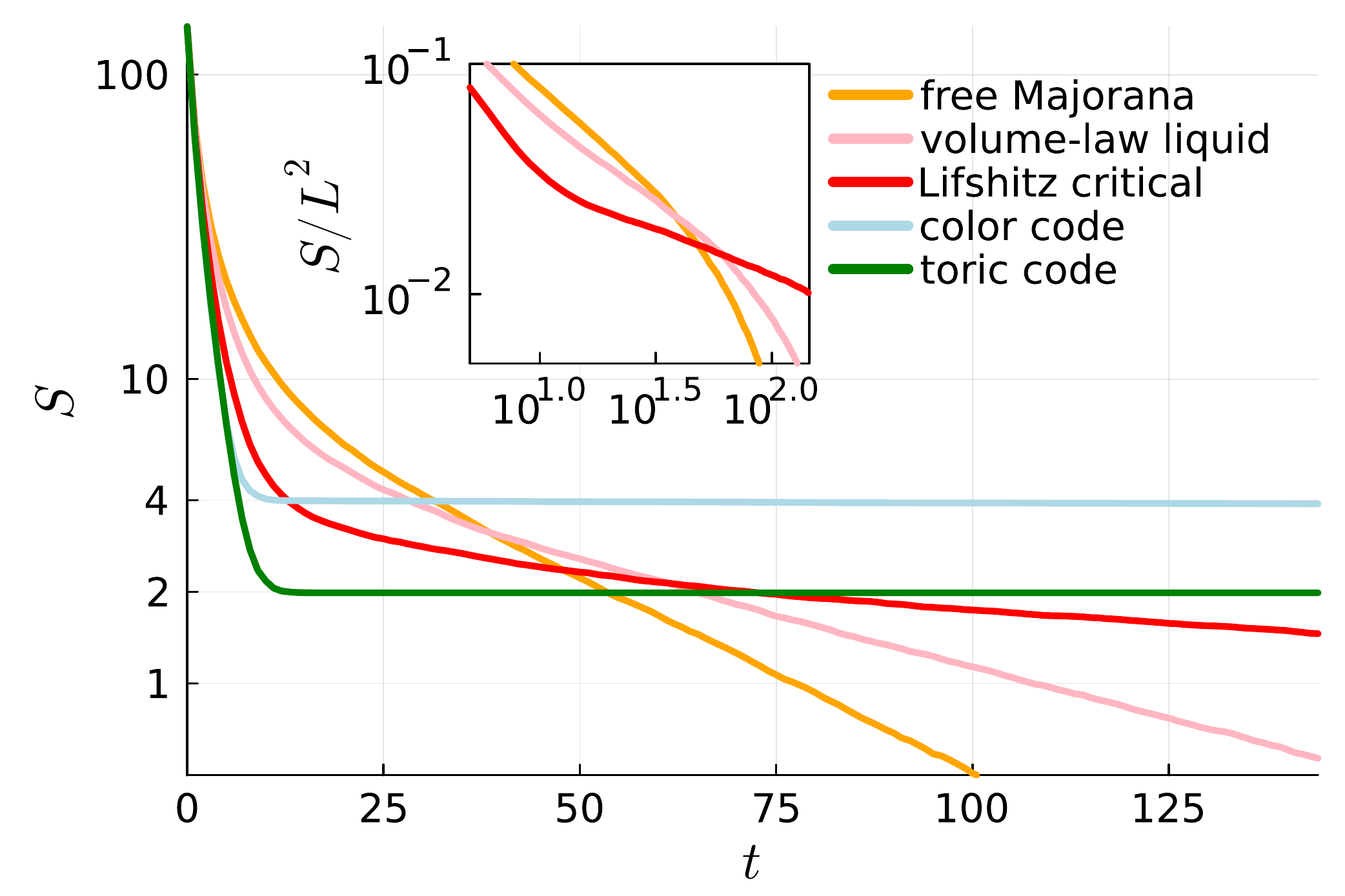} 
   \caption{
   {\bf Purification dynamics of entropy} starting from maximally mixed state in the flux-free space.  $L=12$, average over 1000 disorder samples. Other than the toric code point for which we take $p_z=0.8, p_x=p_y=0.1$, all other cases are adopted along the isotropic line $p_x=p_y=p_z$, by taking $p=0, 0.25, 0.683, 0.8$ for free Majorana, volume-law liquid, Lifshitz critical, and color code, respectively. Do not confuse this thermal entropy, characterizing the mixed state, with the entanglement entropy for pure states we discussed previously. Inset shows the same data on a double log scale.}
   \label{fig:purify}
\end{figure}

\begin{figure}[H] 
   \centering
   \includegraphics[width=\columnwidth]{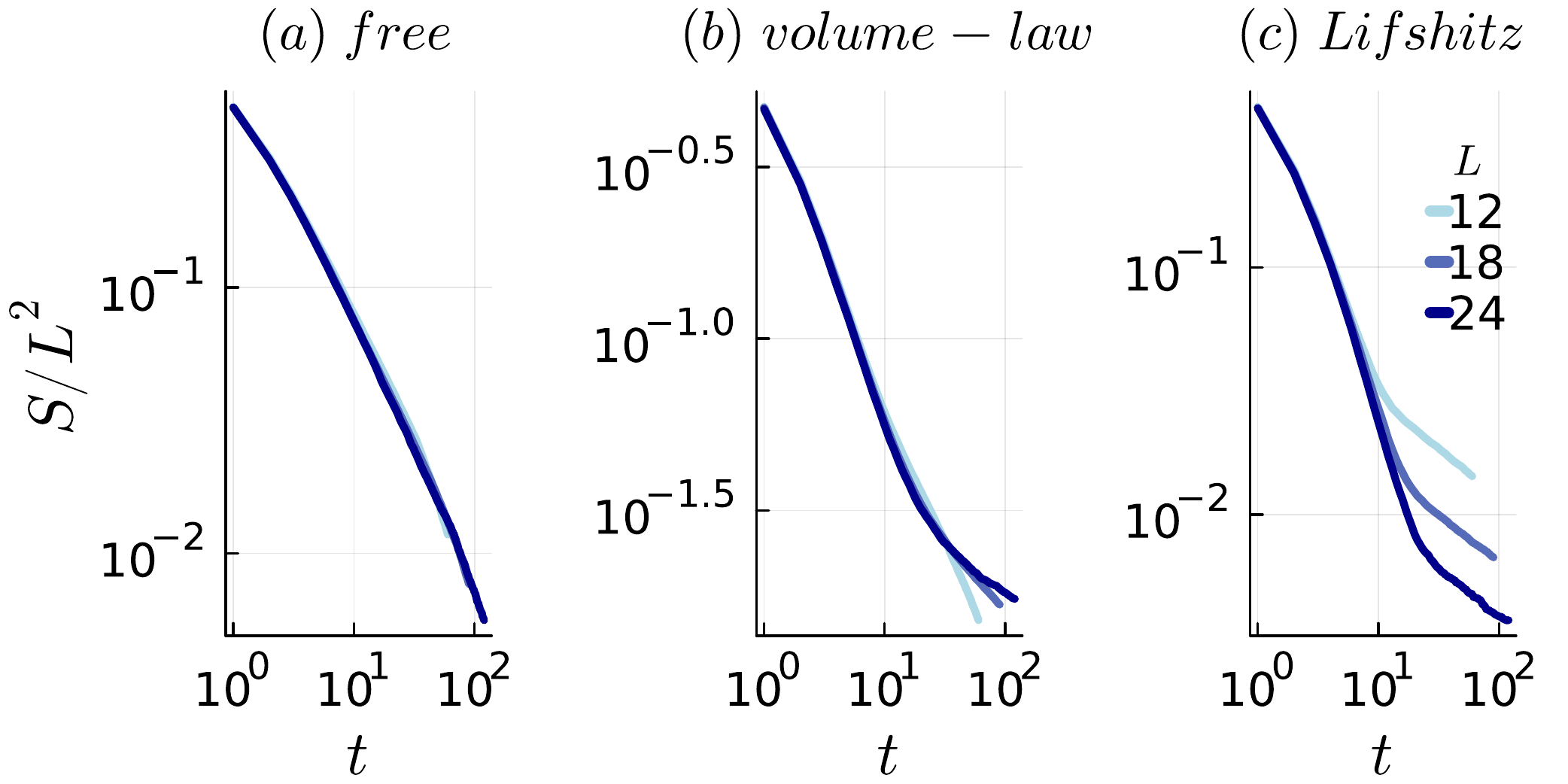} 
   \caption{
  {\bf System-size dependence of purification dynamics} of  the entropy for the gapless states ($p_x=p_y=p_z$, $p=0, 0.25, 0.683$). Data is averaged over 100, 500, 1000 samples for $L=24, 18, 12$, respectively.}
   \label{fig:purifyLvary}
\end{figure}

Both the topological code and the volume-law phase can serve as quantum error correcting codes. Starting with an initial mixed state, which might encode some information, the dynamics leads to a purification of the mixed state which implies a corruption of the code space and the information stored in it~\cite{Altman2020qec, Huse2020qec}. Here we perform an exploratory calculation, for small system sizes, of the purification dynamics of our quantum circuit starting from a maximally mixed state in the flux-free space
$	
	\rho \propto \prod_{q=1}^{L^2-1} \left(1+W_q\right) \,,
$ 
where $q$ denotes a hexagon plaquette. 
Then the purification is purely dominated by the Majorana fermions. 
As shown in Fig.~\ref{fig:purify}, the toric code phase (at $p=0, p_z=0.8, p_x=p_y=0.1$) and the color code phase (at $p=0.8, p_z=p_x=p_y$) both show exponentially long life-times, i.e.\ robust topological code space, exhibiting 2-bit and 4-bit logical memory, respectively. 
The gapless states, in contrast, do not possess such long-lived plateaus. For these, we pick three representative points along the isotropic line $p_x=p_y=p_z$ with $p=0, 0.25, 0.683$, where the finite size dependence is shown in Fig.~\ref{fig:purifyLvary}. They all exhibit similar size-independent power-law decay at the beginning stage, but show very distinct deviation pattern away from that at late-times. 
(i) The free Majorana state exhibits size independent power-law decay $S \sim 1/t$, which was explained in the compact loop model by the L\'evy flight of free Majorana pairs~\cite{Vijay22}. 
(ii) 
The volume-law interacting Majorana liquid  deviates from the power-law by slowing down at late-times, which we attribute to the information scrambling that resists against purification and corruption of logical memories. 
(iii)
The quantum Lifshitz critical point also shows similar slowdown at late-times.

\end{document}